\documentclass{article}

    \usepackage[final,nonatbib]{tccml_neurips_2020}

\usepackage[utf8]{inputenc} %
\usepackage[T1]{fontenc}    %
\usepackage{hyperref}       %
\usepackage{url}            %
\usepackage{booktabs}       %
\usepackage{amsfonts}       %
\usepackage{nicefrac}       %
\usepackage{microtype}      %
\usepackage{cleveref}
\usepackage{siunitx}
\usepackage{etoolbox}
\usepackage{svg}
\usepackage{diagbox}
\usepackage{subcaption}
\usepackage{authblk}

\robustify\bfseries

\newcommand{\model}{ConvMOS}

\title{Deep Learning for Climate Model Output Statistics}

\author[1]{Michael Steininger}
\author[2]{Daniel Abel}
\author[2]{Katrin Ziegler}
\author[1]{Anna Krause}
\author[2]{Heiko Paeth}
\author[1]{Andreas Hotho}
\affil[1]{Chair of Computer Science X (Data Science), University of W\"urzburg \authorcr\{\tt steininger, anna.krause, hotho\}@informatik.uni-wuerzburg.de}
\affil[2]{Chair of Physical Geography, University of W\"urzburg \authorcr\{\tt daniel.abel, katrin.ziegler, heiko.paeth\}@uni-wuerzburg.de}

\begin{document}

\maketitle

\begin{abstract}
  Climate models are an important tool for the assessment of prospective climate change effects but they suffer from systematic and representation errors, %
  especially for precipitation.
  Model output statistics (MOS) reduce these errors by fitting the model output to observational data with machine learning.
  In this work, we explore the feasibility and potential of deep learning with convolutional neural networks (CNNs) for MOS.
  We propose the CNN architecture \model{} specifically designed for reducing errors in climate model outputs and apply it to the %
  climate model REMO.
  Our results show a considerable reduction of errors and mostly improved performance compared to three commonly used MOS approaches.
\end{abstract}

\section{Introduction}
An important source of information for the prospective effects of climate change are numerical climate models such as general circulation models (GCMs) and regional climate models (RCMs).
However, these climate models often exhibit systematic errors and deficiencies in representations of climate processes which limit the quality of the resulting projections.
This problem is especially pronounced for precipitation.
It is therefore common to apply model output statistics (MOS), which are statistical post-processing techniques to reduce these errors.
MOS approaches correct the modeled precipitation to correspond more closely to observational data.
This allows us to study future climate conditions and the effects of climate change more accurately especially at a local scale~\cite{paeth2011postprocessing}.

There are two general approaches to MOS -- distribution-wise MOS and event-wise MOS.
Distribution-wise MOS corrects the distribution of the simulated variable by mapping means and other distribution characteristics to the observational distribution.
Event-wise MOS directly links the simulated and observed time series through statistical models, which generally performs better than distribution-wise MOS~\cite{eden2014downscaling}.
We therefore consider event-wise MOS in this work.

A number of approaches to event-wise MOS have been used in previous work.
A very simple approach is local scaling where an individual Linear Regression is fitted per location of interest, which has shown to work reasonably well~\cite{eden2014downscaling}.
Other works propose non-local MOS approaches, where for each location the MOS is aware of the climatic conditions at nearby locations.
This can lead to a large number of predictors for the MOS, which is why dimensionality reduction techniques, e. g. principal component analysis (PCA), are often applied~\cite{paeth2011postprocessing,eden2014downscaling,sa2017projection,noor2019non}.
Non-local MOS has been done with a range of machine learning models namely Linear Regression~\cite{paeth2011postprocessing,eden2014downscaling}, Random Forests (RFs)~\cite{sa2017projection,noor2019non}, Support Vector Machines (SVMs)~\cite{sa2017projection,pour2018model,ahmed2019modeling}, and Multilayer Perceptrons (MLPs)~\cite{moghim2017bias}.

While these methods have proven to be effective, we believe that there is considerable potential in exploring the use of advanced deep learning techniques.
Especially convolutional neural networks (CNNs)~\cite{lecun1998gradient} have shown proficiency in tasks with geospatial data~\cite{shi2017deep,steininger2020maplur}, which indicates potential for novel non-local MOS with this type of neural network.
We believe that their ability to learn spatial patterns is well suited for reducing systematic errors in climate models.
It is therefore promising to assess how this can be used for MOS and whether there is potential for performance improvements.

In this work, we examine the feasibility and potential of convolutional deep learning models as MOS.
Thus, we propose the CNN architecture \model{} specifically designed for climate MOS and apply it to correcting simulated precipitation of the RCM REMO~\cite{majewski1991europa,jacob2001note,jacob2001comprehensive}.
Our results show that \model{} can reduce errors considerably, providing mostly better performance than three commonly used MOS approaches.
This suggests that our proposed approach is feasible and promising.

\section{Dataset}
\label{sec:dataset}

\paragraph{Model Data} For our study we use daily data of the hydrostatic version of the RCM REMO (version REMO2015)~\cite{majewski1991europa,jacob2001note,jacob2001comprehensive} for the period 2000 to 2015.
Our study area has been defined over an extended German region with \SI{0.11}{\degree} $\times$ \SI{0.11}{\degree} resolution covering the area from \SIrange{-1.43}{22.22}{\degree}~E and \SIrange{42.77}{57.06}{\degree}~N (GER-11).
We use the following \num{22} MOS predictors from REMO:
Daily mean, minimum and maximum temperature \SI{2}{\meter} above surface~[\si{\kelvin}], u- and v-wind \SI{10}{\meter} above surface~[\si{\meter\per\second}], sea level pressure~[\si{\pascal}] and total precipitation (convective + large scale + snowfall)~[\si{\milli\meter}].
Further, the temperature [\si{\kelvin}], geopotential height~[\si{\meter}], and specific humidity~[\si{\kilogram\per\kilogram}] in the pressure levels of \numlist{100;200;500;850;950}~\si{\hecto\pascal} are used.
For elevation information the dataset GTOPO (\SI{0.009}{\degree} $\times$ \SI{0.009}{\degree}) \cite{daac1996gtopo, gesch99} is used by REMO, which we also employ as another predictor for \model{}.
More specifics about the climate model can be found in~\Cref{sec:climate_model_data}.

\paragraph{Observational Data} For observational data we use the gridded dataset E-OBS~\cite{haylock08} version 19.0e which is based on an ensemble of interpolated station data~\cite{cornes2018ensemble}.
Since the station density varies in space and time, the interpolation of the station data has some uncertainty~\cite{cornes2018ensemble}. %
Amongst other variables E-OBS provides daily precipitation sums at a \SI{0.1}{\degree} resolution, which is our predictand.
The grids of the model and observational data are interpolated bilinearly to the same \SI{0.11}{\degree} grid~\cite{schulzweida19}. %

\section{Deep Learning for Climate Model Output Statistics}
To explore the use of deep learning and CNNs as MOS we propose the architecture \model{}.

\paragraph{Idea} The basic idea of \model{} stems from two potential sources of error in climate models:
First, specific location errors which typically stem from poor grid point representation of topography~\cite{paeth2011postprocessing,eden2014downscaling}. %
Second, systematic errors originating from parameterization, which replaces too complex or too small-scale processes with simplified variants.
For precipitation, cloud and rainfall formation is based on parameterization, leading to an overestimation over land~\cite{paeth2011postprocessing}.

To efficiently reduce both types of errors we combine per-location model parameters, which can learn the characteristics of a specific location, and global model parameters, which can learn spatial precipitation patterns to efficiently help reduce systematic errors in climate models.
Thus, we define two module types: Local network and global network.

\paragraph{Local Network} The local network module contains individual model parameters for each location in the study area, allowing it to reduce specific local errors. %
It is implemented with a linearly activated 1D CNN where the input at each time is first reshaped so that it has the dimensions $(\text{height} * \text{width}, \text{predictors})$ instead of $(\text{predictors}, \text{height}, \text{width})$.
In conjunction with setting the kernel size equal to the number of predictors, this allows us to group the convolution for each input channel (i.e. each location) so that each location is convolved with its own set of filters for all predictors.
Thus each location has its own model parameters, in which location characteristics can be encoded.
This module is not provided with elevation data as it would be static across all times for each location. %
The output of the local network is a grid with precipitation residuals for each location.

\paragraph{Global Network} The global network learns spatial patterns in precipitation and other predictors.
This can be done efficiently with CNNs~\cite{vandal2017deepsd}.
The module contains a 2D CNN with \num{4} layers which learns useful filters for the reduction of systematic errors across the study area using all predictors.
Starting from the first layer, the layers have \num{4}, \num{8}, \num{16}, and \num{1} filters and kernel sizes of \num{9}, \num{1}, \num{5}, and \num{3} respectively.
Each convolutional layer has its padding parameter set to half its kernel size (rounded down to the nearest whole number) which leads to the output of each layer having the same width and height as its input.
All layers use the ReLU~\cite{nair2010rectified} activation function, a stride of \num{1}, and a dilation of \num{1}.
As with the local network, this module also outputs a grid of precipitation residuals.

\begin{figure}
    \centering
    \includegraphics[width=0.95\textwidth]{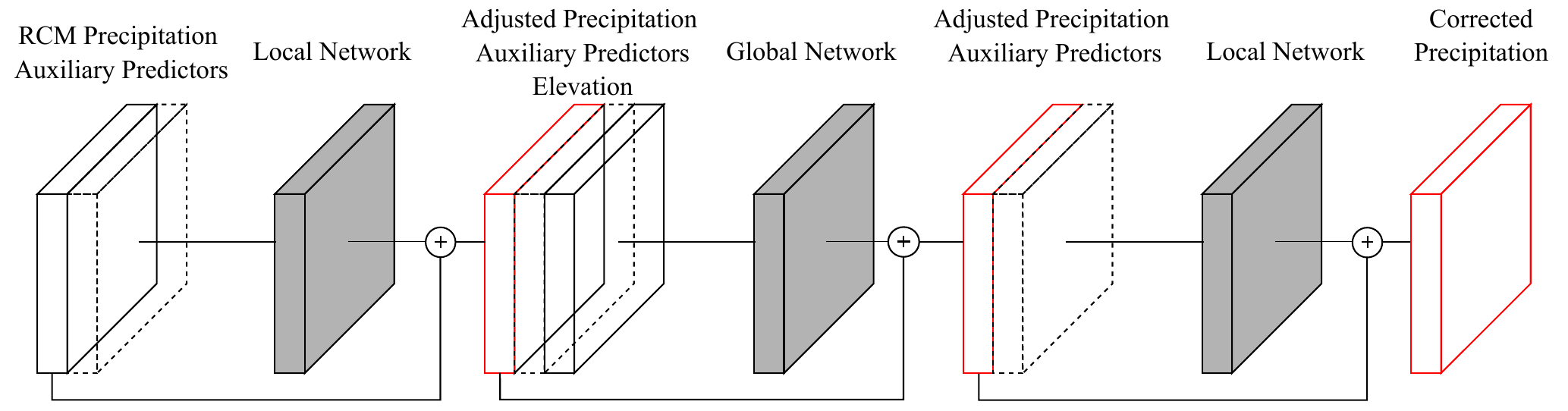}
    \caption{Architecture of \model{}.}
    \label{fig:model_architecture}
\end{figure}

\paragraph{Architecture} The architecture is depicted in~\Cref{fig:model_architecture}.
It expects a 3D input with dimensions $(\text{predictors}, \text{height}, \text{width})$ for each time step.
The data is sequentially passed through three modules (depicted in gray) where each module adjusts the precipitation input with the goal of reducing the error.
The architecture employs so called ``shortcut connections'' for each module where each module's output is added to its precipitation input, which eases training for neural networks~\cite{he2016deep}.
We first apply a local network to correct any specific local errors.
Thereafter, the global network aims to reduce any remaining systematic errors across the study area.
Finally, another local network makes sure that the systematic corrections of the global network are not introducing new local errors.
The training procedure is described in~\Cref{sec:training_details}.

\section{Experiment}
To evaluate \model{} we apply it to the data described in~\Cref{sec:dataset}. %
We also apply three other commonly used MOS approaches, a local Linear Regression, a non-local Principal Component Regression approach and a non-local RF method, for comparison.

\paragraph{Local Linear Regression (Lin)} For each cell in the study area a separate Linear Regression is fitted where the predictor is the simulated precipitation and the predictand is the observed precipitation. %
This approach is local in that each Linear Regression is unaware of conditions in nearby cells~\cite{eden2014downscaling}.

\paragraph{Non-local Principal Component Regression (NL PCR)} Instead of only using the large-scale conditions at a specific location for a Linear Regression, we provide all available predictors at each nearby location which is at most $\pm 5$ cells away in either direction on the grid.
To reduce the dimensionality of the predictors, we apply a supervised PCA~\cite{bair2006prediction}, which is explained in~\Cref{sec:spca}~\cite{eden2014downscaling}.

\paragraph{Non-local Random Forest (NL RF)} For the non-local Random Forest MOS approach we provide all available predictors of each location $\pm 5$ cells away, as with the non-local PC regression approach.
Following~\cite{sa2017projection} and~\cite{noor2019non} we also apply a supervised PCA (see~\Cref{sec:spca}).
Each location is fitted with its own RF.
Hyperparameters are optimized at each location individually (see~\Cref{sec:rf_hpopt}).

\paragraph{Experimental Setup} We split the \num{16} years of daily data into a training (2000--2009), a validation (2010), and a test set (2011--2015).
All predictors are standardized based on the training set so that they have a mean of zero and a standard deviation of one.
We tried different hyperparameters for our architecture and selected the ones presented in this work based on the validation set performance.
All results reported in the following are based on the unseen test set after fitting the MOS on the training set.
For evaluation we use a number of common MOS metrics, namely RMSE, Pearson Correlation, Skill Score~\cite{perkins2007evaluation}, $\text{R}^2$, and bias to assess different aspects of MOS performance.
\model{} is trained \num{10} times since its fitting method is, in contrast to the linear comparison methods, non-deterministic.
This results in slight performance differences for each fitted instance.
Despite its inherent randomness, the RF method is only calculated once since this computation already took over four days for our study area with 15 CPU cores in parallel.

\renewrobustcmd{\bfseries}{\fontseries{b}\selectfont}
\renewcommand{\pm}{\mathbin{\mbox{\unboldmath$\mathchar"2206$}}}

\begin{table}[]
    \centering
    \caption{Experimental results. Mean metrics on the test set for all study area locations available in observational data. All means and standard deviations are rounded to two decimal places. Correlation mean is calculated with Fisher's z-transformation~\cite{silver1987averaging}.}
    \label{tab:results}
    
    \sisetup{separate-uncertainty,detect-weight=true,detect-family=true,detect-inline-weight=math,mode=text}
    \begin{tabular}{@{}
        l
        S[table-format=1.2,table-figures-uncertainty=1]
        S[table-format=1.2,table-figures-uncertainty=1]
        S[table-format=1.2,table-figures-uncertainty=1]
        S[table-format=1.2,table-figures-uncertainty=1]
        S[table-format=1.2,table-figures-uncertainty=1]
        S[table-format=1.2,table-figures-uncertainty=1]
        @{}}
    \toprule
    \diagbox[trim=l]{MOS}{Metric} &   {RMSE} &   {Corr.} &   {Skill} &   {$\text{R}^2$} &  {Bias}  \\
    \midrule
    None & 5.32 & 0.49 & \bfseries 0.93 & -28.24 & 0.31 \\
    \midrule
    Lin &  3.77 &  0.49 & \bfseries 0.93 &  0.23 &  0.03 \\
    NL PCR & 3.37 & 0.62 & 0.92 & 0.36 & \bfseries 0.02  \\
    NL RF & 3.39 & 0.61 & 0.81 & 0.36 & 0.03 \\
    \model{} & \bfseries 2.99(1) & \bfseries 0.72(0) & 0.92(0) & \bfseries 0.49(1) & -0.10(6) \\  %
    \bottomrule
    \end{tabular}

\end{table}
    
\paragraph{Results} \Cref{tab:results} shows the mean metrics on the test set for all study area locations available in observational data (i.e. land points). %
All MOS approaches improve all metrics considerably when compared to applying no MOS, except for the skill score.
This means that the precipitation distribution of REMO is already rather close to that of the observations with a skill score of \num{0.93} and can barely be improved by the MOS methods.
\model{} is showing the best performance of all tested MOS approaches for the metrics RMSE, correlation, and $\text{R}^2$.
This indicates that our approach is able to estimate precipitation more accurately than all considered comparison methods.
The skill score is very close but still reduced slightly by \num{0.01} compared to the best value.
\model{} shows less bias than REMO but it seems to have a tendency to underestimate precipitation.
The other approaches to tend to overestimate, but to a lesser extent.
\model{} is also showing rather stable performance as can be seen on the standard deviations in \Cref{tab:results} despite its non-deterministic fitting procedure.
We also ran this experiment with precipitation as the only climate predictor as some prior work has done~\cite{eden2014downscaling,sa2017projection,noor2019non,ahmed2019modeling} but found all methods to perform worse without additional predictors.

\begin{figure}
    \centering
    \begin{subfigure}{0.455\textwidth}
        \centering
        \includegraphics[width=\textwidth]{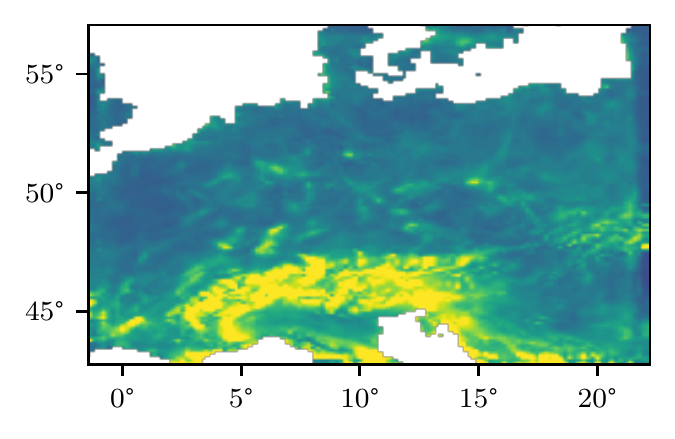}
        \caption{REMO raw}
    \end{subfigure}%
    ~ 
    \begin{subfigure}{0.505\textwidth}
        ~
        \centering
        \includegraphics[width=\textwidth]{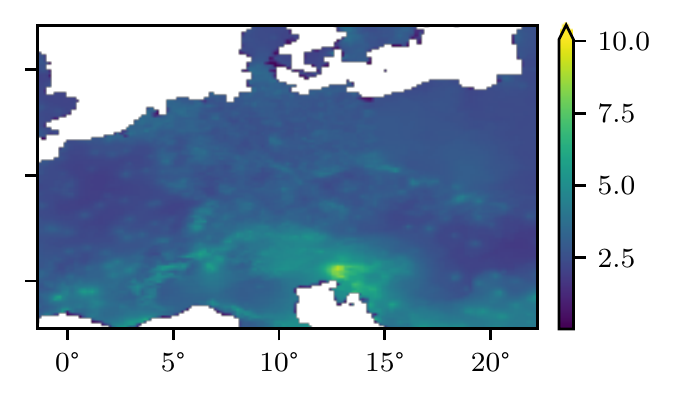} %
        \caption{\model{}}
    \end{subfigure}
    \caption{RMSE of precipitation in \si{\milli\meter} for the test set across the study area. Note that there are some cells in REMO raw with considerably higher RMSE than \SI{10}{\milli\meter} but we limited the colorbar's extent for better visibility of the general performance.}
    \label{fig:rmse_maps}
\end{figure}

\Cref{fig:rmse_maps} visualizes RMSEs for all locations with observational data across the study area for the raw REMO output and \model{}. %
We can see that our approach reduces error across most locations.
Especially the precipitation in the Alps and other mountainous regions is improved considerably.

\section{Conclusion}
In this work we explored the feasibility and possibilities of deep learning MOS.
To this end, we proposed the CNN-based \model{} architecture specifically designed to reduce errors in climate model outputs, which we applied to the RCM REMO.
All in all, the initial results for \model{} seem promising.
Our MOS approach is able to improve the daily precipitation data considerably.
Improvements in MOS allow for more accurate climate data especially at high spatial resolutions.
While our approach mostly provides better performance than the other standard approaches considered here, additional comparisons have to be made in the future with other MOS techniques and data from different climate models.

\begin{ack}
This research was conducted in the BigData@Geo project supported by the European Regional Development Fund (ERDF).
\end{ack}

\bibliographystyle{unsrt}
\bibliography{references}

\newpage
\appendix

\section{Climate Model Data}
\label{sec:climate_model_data}

For our study we use daily data of the hydrostatic version of the RCM REMO (version REMO2015)~\cite{majewski1991europa,jacob2001note,jacob2001comprehensive} for the period 2000 to 2015.
REMO is based on the Europa Modell \cite{majewski1991europa} with the model of the GCM ECHAM4~\cite{roeckner96} with some improvements implemented since then (e.g. \cite{hagemann02, semmler02, kotlarski07, pietikainen12}).
The reanalysis ERA-Interim ($0.75^{\circ} \times 0.75^{\circ}$) \cite{dee11, berrisford11} is used as forcing data, providing the lateral boundary conditions.
The atmosphere’s vertical resolution is represented by 27 hybrid levels with increasing distance to the top of the atmosphere.
In lower levels they follow the topography~\cite{teichmann09}.
As mentioned in the main paper the dataset GTOPO ($0.009^{\circ} \times 0.009^{\circ}$) \cite{daac1996gtopo, gesch99} is used by REMO for elevation information.
Both model and observational data for the MOS methods is provided at \SI{0.11}{\degree} resolution.
The data is arranged on a 2D grid with $121 \times 121$ cells or locations.

\section{\model{} Training Details}
\label{sec:training_details}

The architecture is fitted with the Adam optimizer~\cite{kingma2014adam}, the mean squared error (MSE) as the loss function and a learning rate of \num{0.001}.
Only errors at locations where observational data is available were incorporated for the MSE.
Training is conducted for at most \num{100000} epochs.
Early stopping is used to stop training when the validation MSE is not improving for more than \num{40} epochs in a row, preventing considerable overfitting~\cite{caruana2001overfitting}.

\section{Random Forest Hyperparameter Optimization}
\label{sec:rf_hpopt}
Each location in our study area has its own RF instance for MOS which uses the RandomForestRegressor from scikit-learn~\cite{scikit-learn}.
Since RF performance depends considerably on its hyperparameters we look for optimal values with a random search.
For each cell we train \num{20} RF instances on the training set with hyperparameter values sampled randomly from the search space shown in~\Cref{tab:rf_hpspace}.
Each instance is evaluated on the validation set.
The RF instance with the best $\text{R}^2$ is then applied on the test set.

\begin{table}[]
    \centering
    \caption{Search space for the RF hyperparameter random search.}
    \label{tab:rf_hpspace}
    
    \begin{tabular}{@{}
        l
        c
        @{}}
    \toprule
    Hyperparameter & Search space  \\
    \midrule
    n\_estimators & 10 -- 2000 \\
    max\_features & 0.01 -- 1.0 \\
    max\_depth & 10 -- 110  \\
    min\_samples\_split & 2 -- 10 \\
    min\_samples\_leaf & 1 -- 10 \\
    bootstrap & True or False \\
    \bottomrule
    \end{tabular}
    
\end{table}

\section{Supervised Principal Component Analysis}
\label{sec:spca}
Like other previous MOS approaches~\cite{paeth2011postprocessing,sa2017projection,noor2019non} we preprocess our predictors for the standard MOS methods to reduce dimensionality and remove potentially unhelpful information.
Like~\cite{sa2017projection} and~\cite{noor2019non} we use supervised PCA~\cite{bair2006prediction}.
First, we select the best predictors based on a univariate regression.
How many of the predictors are retained is set according to a grid search with our validation data.
In this search we try all values between only choosing the single best predictor and using the \num{30} best predictors.
Then, PCA reduces the dimensionality of these predictors, keeping the first components that explain \SI{95}{\percent} of the variance~\cite{sa2017projection}.

\end{document}